\documentclass[11pt]{article}  
\usepackage{epsfig,graphicx,graphics,amsmath}  
\input epsf  
  
\newcommand{\beq}{\begin{equation}}  
\newcommand{\eeq}{\end{equation}}  
\newcommand{\ben}{\begin{enumerate}}
\newcommand{\een}{\end{enumerate}}
\newcommand{\bitem}{\begin{itemize}}
\newcommand{\eitem}{\end{itemize}}
\newcommand{\bfig}{\begin{figure}}
\newcommand{\efig}{\end{figure}}
\newcommand{\bcen}{\begin{center}}
\newcommand{\ecen}{\end{center}}

\newcommand{\delete}[1]{}

\title{Galactic Rotation Described with Bulge+Disk Gravitational Models}
\author{C. F. Gallo and James Q. Feng
\\ \\ Superconix Inc, 2440 Lisbon Ave, Lake Elmo  MN  55042}

\begin{document}
\pagestyle{myheadings}

\maketitle

\bcen
\Large{\bf Abstract}
\ecen


Observations reveal that mature spiral galaxies consist of stars, 
gases and plasma approximately distributed in a thin disk of circular shape, 
usually with a central bulge. 
The rotation velocities quickly increase
from the galactic center and then achieve a constant velocity 
from the core to the periphery.  
The basic dynamic behavior of a mature spiral galaxy, 
such as the Milky Way, 
is well described by simple models  
balancing Newtonian gravitational forces against 
the centrifugal forces associated with a rotating thin axisymmetric disk.
In this research, we investigate the effects of adding  
central bulges to thin disk gravitational models. 
   
Even with the addition of substantial central bulges, all the critical essential features of our thin disk gravitational models are preserved. 
(1) Balancing Newtonian gravitational and centrifugal forces 
at every point within the disk yields computed radial mass distributions that describe the measured rotation velocity profiles of mature spiral galaxies successfully. 
(2) There is no need for gravity deviations or ``massive peripheral spherical halos of mysterious Dark Matter''. 
(3) The calculated total galactic masses are in good agreement with star count data. 
(4) The addition of central bulges increases the calculated total galactic masses, possibly more consistent with the presence of galactic gases, dust, grains, lumps, planets and plasma in addition to stars. 
(5) Compared with the light distribution, our mass distributions within the disk are larger out toward the galactic periphery which is cooler with lower opactiy/emissivity (and thus darker). This is apparent from edge-on views of galaxies which display a dark disk-line against a much brighter galactic halo.     


\section{Introduction} 

\subsection{Observational Knowledge of Radial Galactic Rotation Profiles}
Telescopic images of mature spiral galaxies reveal most of the stars, 
gas and plasma reside in an approximately circular disk 
that is very thin compared with its radius,
often with the presence of a central bulge.
The data on galactic rotational velocity profiles 
(Refs.\cite{Rubin1}-\cite{deBlockMH}) of mature spiral galaxies are 
characterized by a rapid increase from the galactic center, 
reaching a nearly constant
velocity from the outer core to the outer periphery.  
These basic measured features may be idealized as  
\beq \label{eq:smooth-V}
V(r) = 1 - e^{-r / R_c} \, , 
\eeq
where $V(r)$ denotes the dimensionless rotational velocity
measured in units of maximum asymptotic rotational velocity $V_0$ and 
$r$ the radial coordinate from the galactic center.
The parameter $R_c$ is a description from the data of 
the various ``core'' radii of different galaxies. 
Typical galactic rotational profiles described by (\ref{eq:smooth-V})
are displayed in Fig 1. As indicated by the measurement data, 
the rotation velocity typically rises linearly from the galactic center 
(as if the local mass was in rigid body rotation), 
and then reach an approximately constant (flat) velocity out to 
the galactic periphery. 

\begin{figure}[htb]
\resizebox{!}{1.25\textwidth}
{\includegraphics{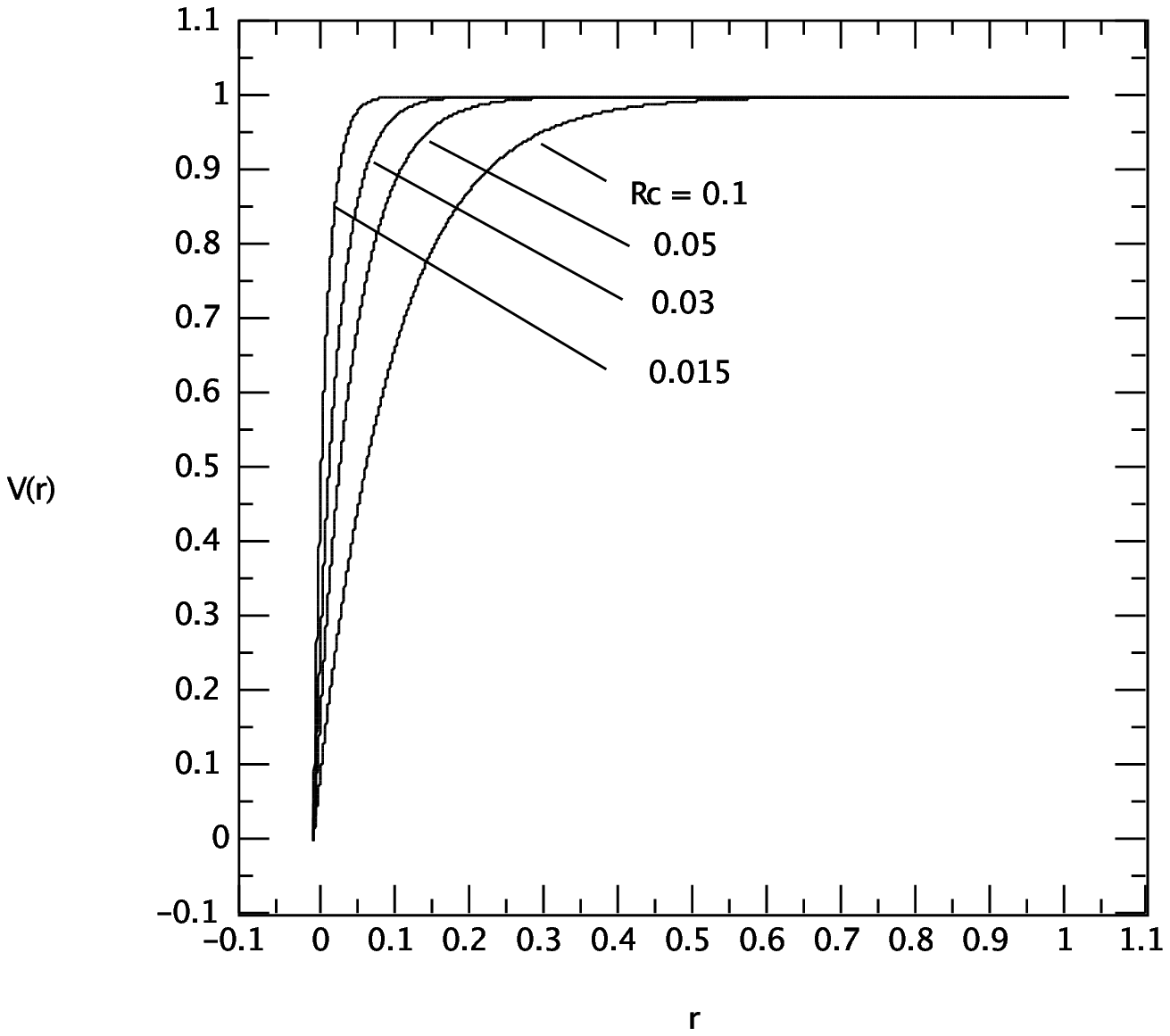}}
\caption{Typical Galactic Rotational Velocity Profiles $V(r)$ idealized (\ref{eq:smooth-V}) from measurements for $R_c = 0.015$, $0.03$, $0.05$, and $0.1$.}
\label{fig:fig1}
\end{figure}

The observed galactic rotation curves (Eq(\ref{eq:smooth-V}) and Fig.1) 
can {\em not} (Ref.\cite{BT})be explained by simply applying the so-called 
{\em orbital velocity law}, derived for a spherically symmetric 
gravitational field applicable to the Keplerian rotation of our solar-planet system
(where most mass is located at the center), 
but {\em not} to galaxies with substantial mass distributed in a disk-like shape.
In fact, the galactic mass distribution calculated by 
the {\em orbital velocity law} applied to these constant (flat) 
galactic rotation curves 
yields an {\em increasing mass density with radius},  
contrary to the measured galactic luminosity curves 
which decrease exponentially with radius.

\subsection{Thin-Disk Gravitational Models with Bulge Added} 
For a thin rotating galactic disk,  
we impose a balance between the 
Newtonian gravitational forces and centrifugal forces at each and every point.  
Because the gravitational field of a thin disk is not spherically symmetric, 
the {\em orbital velocity law} is not applicable.
As illustrated by Feng \& Gallo \cite{FengGallo1} \cite{FengGallo2}, 
an axisymmetric thin disk gravitational model successfully   
describes the basic rotational dynamics of mature spiral galaxies
with a mass density decreasing from the center to periphery. 
And the calculated total galactic masses are in good agreement with star count data.

For simplicity, the observed central bulge in mature spiral galaxies was not considered in previous idealized thin disk gravitational models by Feng \& Gallo \cite{FengGallo1} \cite{FengGallo2}.  
In this paper, consistent with observations, we add the gravitational effects 
of central spherical bulges to thin disk models.  
We do not address the mechanism(s) maintaining the spherical shape against gravitational and centrifugal forces in this publication \cite{GalloVortex}. 
As observed, the central spherical bulge is implicitly assumed to rotate at the same radial speeds as the disk (in cylindrical coordinates), 
but this feature is not explicitly addressed because it does not affect the computational results. 
Only the gravitational effects of this assumed central bulge and the disk 
are computed in solving for the rotating disk mass distributions.

In detail, it is assumed the bulge has a spherically symmetric mass density decreasing with 
radial distance (in spherical coordinates) via a Gaussian function
$e^{-\beta \, r^2}$ where $\beta$ is a positive adjustable parameter.
This Gaussian function is convenient since changing one parameter 
$\beta$ allows us to vary the size of the spherical bulge relative to the disk to examine the effects of the bulge size on galactic disk rotation. 
Our final generic results are not sensitive to the details of this Gaussian assumption. 
Our model of the entire galaxy consists of a variable superposition of two components:
an axisymmetric thin disk and a spherically symmetric central bulge.
Both the size and mass of the bulge relative to the disk are independently varied
to examine the effects of the bulge on galactic disk rotation.

\section{Governing Equations} 
Similar to the treatment of 
Feng \& Gallo \cite{FengGallo1} \cite{FengGallo2}, 
the equation of force balance in an axisymmetric thin disk 
including a spherically symmetric central bulge is written as
\beq \label{eq:force-balance0}
\int_0^1 \left[\int_0^{2 \pi} 
\frac{(\hat{r} \cos \phi - r) d\phi}
{(\hat{r}^2 + r^2 - 2 \hat{r} r \cos \phi)^{3/2}}\right] 
\rho(\hat{r}) h \hat{r} d\hat{r}
- \frac{M_b}{r^2}
\frac{\int_0^r e^{- \beta \hat{r}^2} \hat{r}^2 d\hat{r}}
{\int_0^1 e^{- \beta \hat{r}^2} \hat{r}^2 d\hat{r}}
+ A \frac{V(r)^2}{r}
 = 0 \, ,
\eeq
where all the variables are made dimensionless by measuring lengths 
(e.g., the radial coordinate $r$, the radial coordinate as 
the variable of integration $\hat{r}$, and the thickness of disk $h$) 
in units of the outermost galactic radius $R_g$, 
mass of bulge ($M_b$) in units of the total galatic mass ($M_g$),
disk mass density ($\rho$) in units of
$M_g / R_g^3$, 
and rotational velocity [$V(r)$] in units of 
the maximum galactic rotational velocity $V_0$.  
The disk thickness $h$ is assumed to be constant and small 
in comparison with the galactic radius $R_g$. 
Actually the physically meaningful quantity here is 
the combined variable $(\rho \, h)$ that represents 
the effective surface mass density on the thin disk.
As long as the disk thickness $h$ is much smaller than $R_g$,
its mathematical effect is inconsequential to the value of $(\rho \, h)$.
The gravitational forces of the finite series of concentric rings 
is described by the first term (double integral) while 
the centrifugal forces are described by the third term.
The second term represents the effects of the 
spherically symmetric central buldge. 

In (\ref{eq:force-balance0}), we call the dimensionless paremeter 
$A$ ``galactic rotation parameter'', 
given by
\beq \label{eq:parameter-A}
A \equiv \frac{V_0^2 \, R_g}{M_g \, G} \, ,
\eeq
where $G$ denotes the gravitational constant, $R_g$ is the 
outermost galactic radius, and $V_0$ is 
the maximum asymptotic rotational velocity. 

As described by Feng \& Gallo \cite{FengGallo1} \cite{FengGallo2}, both  
$\rho(r)$ and $A$ can be determined from a given $V(r)$ 
and a given (but varied) value of $M_b$ by 
solving an equation system including (\ref{eq:force-balance0}) and 
a conservation constraint for constant
total mass of the galaxy $M_g$, e.g.,   
\beq \label{eq:mass-conservation}
2 \pi \int_0^1 \rho(\hat{r}) h \hat{r} d\hat{r} = 1 - M_b \, .  
\eeq

Assuming a bell-shape mass density distribution for the spherically
symmetric bulge described by a Gaussian function,
the bulge mass density is given by
\beq \label{eq:bulge-mass-density}
\rho_b(r) = \frac{M_b}{4 \pi \int_0^1 e^{- \beta \hat{r}^2} \hat{r}^2 d\hat{r}}
e^{- \beta r^2} \, ,
\eeq
where $\beta$ is a positive adjustable parameter.
The bulge mass density is assumed to end at the galaxy rim ($r = 1$),
so that
\[
4 \pi \int_0^1 \rho_b(\hat{r}) \hat{r}^2 d\hat{r} = M_b \, .
\]

\begin{figure}[htb]
\resizebox{!}{1.25\textwidth}
{\includegraphics{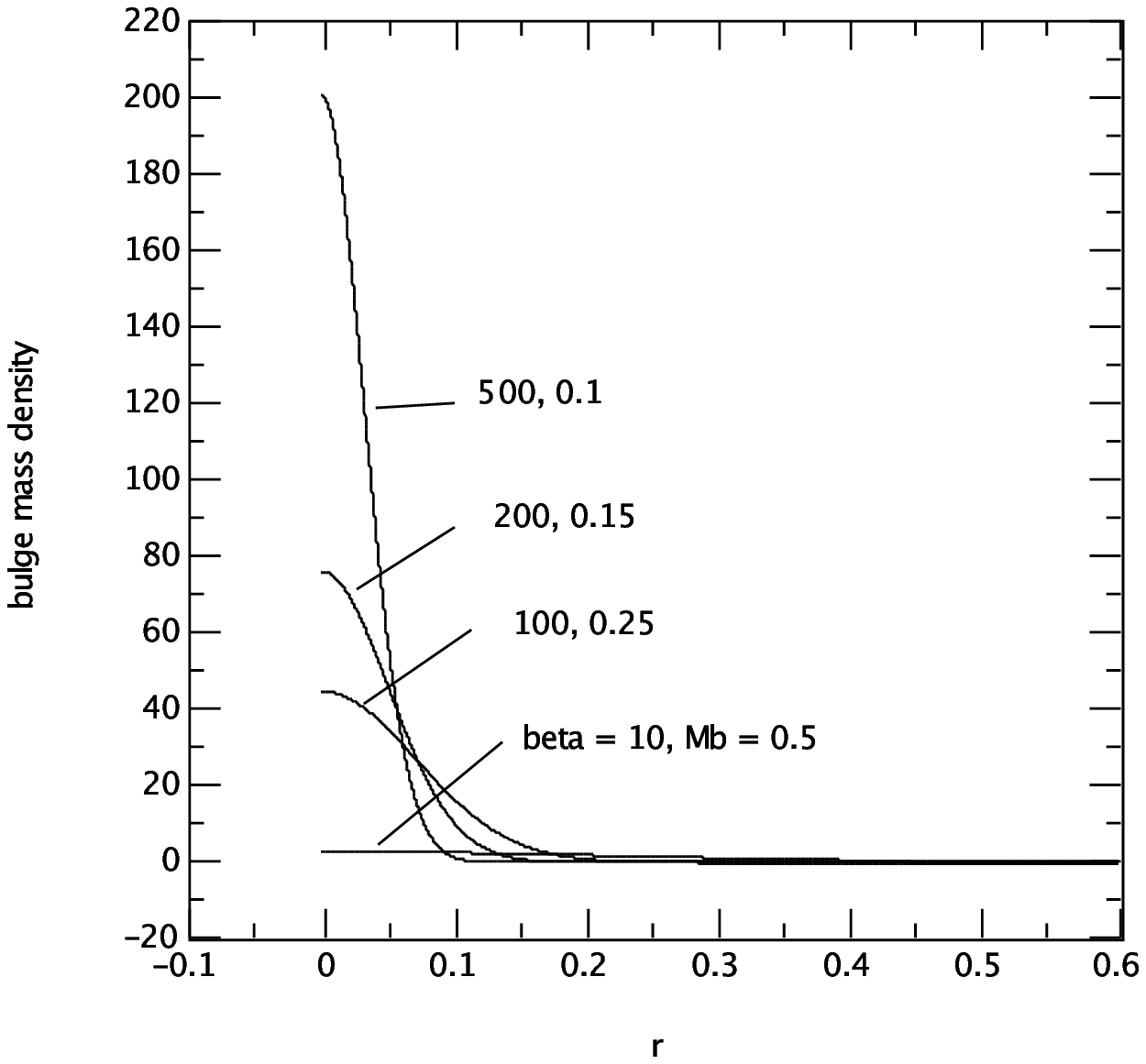}}
\caption{Bulge mass density distributions $\rho_b(r)$ from (\ref{eq:bulge-mass-density}) are displayed for reasonable bulge parameters we have examined ($\beta = 10$, $100$, $200$, and $500$  
with $M_b = 0.5$, $0.25$, $0.15$, and $0.1$).}
\label{fig:rho_b}
\end{figure}

Figure \ref{fig:rho_b} illustrates 
several bulge mass density distributions $\rho_b(r)$ at
various values of $\beta$ and $M_b$. 
Note the bulge mass density $\rho_b(r)$ is
spherically symmetric, whereas the disk mass density $\rho(r)$ 
is only axisymmetric and $r$ denotes the radial disk coordinate
in the cylindrical coordinate system 
used in the present computations. 

\section{Computational Techniques} 
To facilitate numerical computation,
we discretize the governing equations 
(\ref{eq:force-balance0}) and (\ref{eq:mass-conservation})
by dividing the one-dimensional problem domain $[0, 1]$
into a finite number of line segments called (linear) elements.
As described by Feng \& Gallo \cite{FengGallo1},
each element covers a subdomain confined by two end nodes,
e.g., element $i$ corresponds to the subdomain $[r_i, r_{i+1}]$
where $r_i$ and $r_{i+1}$ are the nodal values of $r$
at nodes $i$ and $i+1$, respectively.  
With each of the $N - 1$ elements mapped onto 
a unit line segment $[0, 1]$ 
in the $\xi$-domain (i.e., the computational domain),
$N$ independent residual equations can be obtained 
from the collocation procedure, i.e.,
\beq \label{eq:force-balance-residual}
\sum_{n = 1}^{N - 1} \int_0^1 \left[
\frac{E(m_i)}{\hat{r}(\xi) - r_i} - \frac{K(m_i)}{\hat{r}(\xi) + r_i}
\right] 
\rho(\xi) h \hat{r}(\xi) \frac{d\hat{r}}{d\xi} d\xi
+ \frac12 \left[A V(r_i)^2
- \frac{M_b}{r}
\frac{\int_0^{r_i} e^{- \beta \hat{r}^2} \hat{r}^2 d\hat{r}}
{\int_0^1 e^{- \beta \hat{r}^2} \hat{r}^2 d\hat{r}}\right]
 = 0 \, ,
\eeq
where $K(m)$ and $E(m)$ denotes the complete elliptic integrals
of  the first kind and second kind, with
\beq \label{eq:mi-def}
m_i(\xi) \equiv \frac{4 \hat{r}(\xi) r_i}{[\hat{r}(\xi) + r_i]^2} \, .
\eeq

The $N$ residual equations (\ref{eq:force-balance-residual}) 
can be used to 
compute either the $N$ nodal values of $V(r_i)$ 
from given distribution of $\rho(r_i)$ or 
the distribution of $\rho(r_i)$ from a given set of $V(r_i)$,
with given values of $A$ and $h$.
Without loss of generality, the value of $h$ is assumed to be $0.01$
as comparable with that observed for the Milky Way galaxy.
If the constraint equation (\ref{eq:mass-conservation}) 
is also used with a discretized form
\beq \label{eq:mass-conservation-residual}
2 \pi \sum_{n = 1}^{N - 1} \int_0^1 
\rho(\xi) h \hat{r}(\xi) \frac{d\hat{r}}{d\xi} d\xi - 1 + M_b = 0 \, ,
\eeq
the value of $A$ can also be determined as part of 
the numerical solution.

These generally applicable equations are conveniently used 
for computing variables, even when analytical formulas are
available for some special cases.
Hence, a unified treatment for all cases is established 
for convenient comparison and analysis.
Moreover, as discussed by Feng \& Gallo \cite{FengGallo1},
imposing a boundary condition at the galactic center $r = 0$
for continuity of derivative of $\rho$,
i.e., in discretized form
\beq \label{eq:rho-1}
\rho(r_1) = \rho(r_2)
\, ,
\eeq
is desirable for obtaining high-quality numerical solutions.

With the adjustable parameters such as $R_c$, $\beta$, and $M_b$ 
specified and mathematical singularities properly treated, linear equations 
(\ref{eq:force-balance-residual})
and
(\ref{eq:mass-conservation-residual})
for $N + 1$ unknowns can be solved with a standard matrix solver,
e.g., by Gauss elimination
\cite{PressTVF}.

\section{Computational Results for Bulge+Disk Models}
Here the effects of bulge parameters on galactic disk mass distributions compatible with measured rotational velocities are explored. 
Attention is focused on the Milky Way galaxy which has a rotation velocity profile (\ref{eq:smooth-V})
closely represented in Figure \ref{fig:fig1} with $R_c = 0.015$ 
(cf. Feng \& Gallo \cite{FengGallo1} \cite{FengGallo2}). 

\subsection{Disk Mass Density Distributions Calculated with Constant $\beta = 100$  
but Various Bulge Masses $M_b$}  
\begin{figure}[htb]
\resizebox{!}{1.25\textwidth}
{\includegraphics{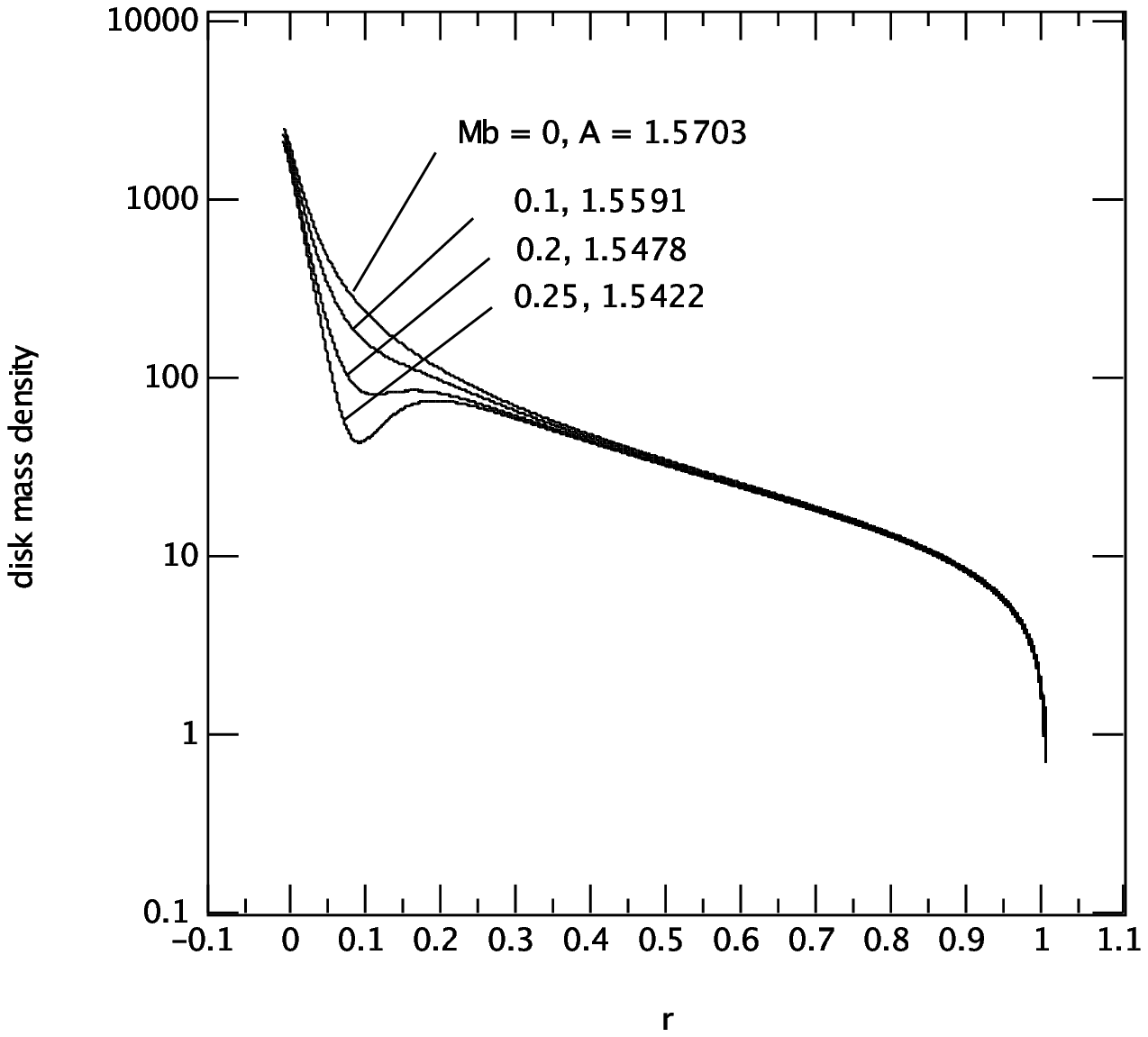}}
\caption{Disk mass density $\rho(r)$ computed 
with $\beta = 100$ and various bulge masses.  
$M_b = 0$, $0.1$, $0.2$, and $0.25$.   
Note the case of $M_b = 0$ represents a thin disk without a bulge. 
All these curves are for the Milky Way galaxy with $R_c = 0.015$. 
In all cases, the appropriate calculated galactic rotation parameters $A$ are indicated. 
In all these bulge + disk cases, the measured galactic rotation profiles are accurately reproduced.}
\label{fig:rho_Rc015_b100}
\end{figure}

Figure \ref{fig:rho_Rc015_b100} shows the disk mass density 
distributions computed with constant $\beta = 100$  but various bulge masses ($M_b = 0$, $0.1$, $0.2$, and $0.25$). 
All these curves are for the Milky Way galaxy with a rotation curve 
with $R_c = 0.015$ according to (\ref{eq:smooth-V}).
Note that $M_b = 0$ represents a thin disk without a bulge. 
With increasing bulge mass $M_b$, 
a localized decrease of disk mass density appears where the 
bulge mass has significant density around the galatic center.
A local minimum develops around $r = 0.1$ for $M_b \ge 0.2$.  
When the value of $M_b$ is further increased, 
the nodal values of  
disk mass density $\rho$
around local minimum may become negative which is physically unacceptable.  
Thus, our computational results demonstrate an upper limit for the bulge mass $M_b$ 
corresponding to a given bulge size as characterized by the value of $\beta$. 
This is consistent with reality. 
In all bulge + disk cases, the measured galactic rotation profiles are accurately reproduced. 

Another noteworthy feature in 
Figure \ref{fig:rho_Rc015_b100} is that
the bulge influence on the disk mass distribution
diminishes beyond $r = 0.2$,
where the bulge reaches its effective edge (cf. Figure \ref{fig:rho_b}).

\subsection{Disk Mass Density Distributions Computed with Various Combinations of Bulge Parameters $M_b$ and $\beta$}   
\begin{figure}[htb]
\resizebox{!}{1.25\textwidth}
{\includegraphics{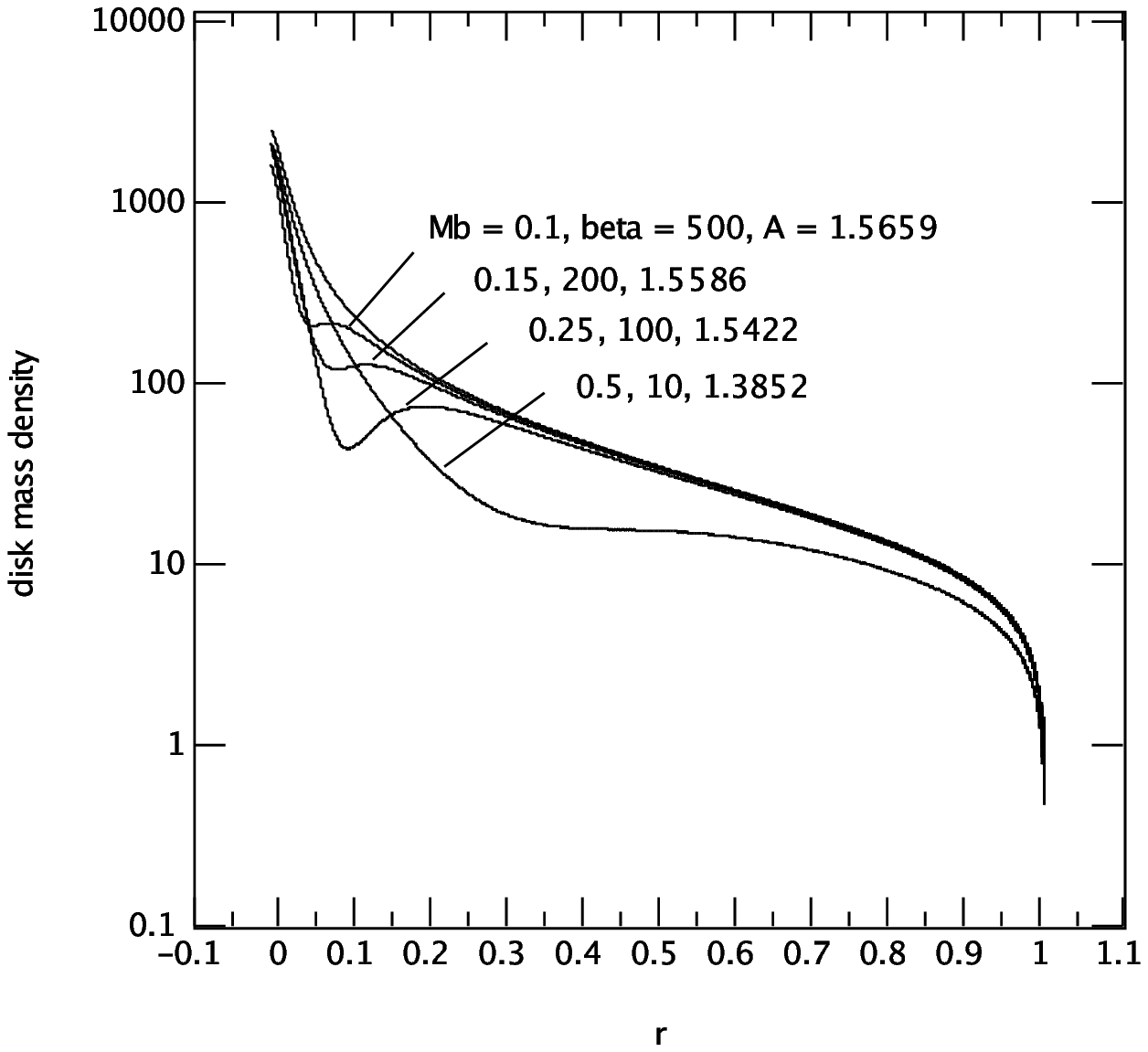}}
\caption{Disk mass density $\rho(r)$ computed for various combinations of bulge parameters. 
For $M_b = 0.1$, $\beta = 500$, and $A= 1.57$.   
For $M_b = 0.15$, $\beta = 200$, and $A= 1.56$. 
For $M_b = 0.25$, $\beta = 100$, and $A= 1.54$.
For $M_b = 0.5$, $\beta = 10$, and $A= 1.39$.  
For comparison, the curve without label is for $M_b = 0$ which represents a disk without a bulge.
In all cases, the appropriate calculated galactic rotation parameters $A$ are indicated. 
The constant value of $R_c = 0.015$ is utilized which represents the Milky Way galaxy.
In all bulge + disk cases, the measured galactic rotation profiles are accurately reproduced.} 
\label{fig:rho_Rc015}
\end{figure}

Figure \ref{fig:rho_Rc015} shows the computed disk mass density 
distributions for various combinations of bulge masses $M_b$ and $\beta$. 
For reference, the unlabled curve is for $M_b = 0$ which represents a thin disk without a bulge.
In all cases, the appropriate calculated galactic rotation parameters $A$ are indicated. 
All these curves are for the Milky Way galaxy with $R_c = 0.015$. 
In all bulge + disk cases, the measured galactic rotation profiles are accurately reproduced. 

\section{Total Galactic Mass}

The measured rotational velocity profiles $V(r)$ 
includes knowledge of maximum rotational velocity $V_0$ and galactic radius $R_g$. 
With the computed value of the
galactic rotation parameter $A$, 
the total galactic mass
$M_g$ can be calculated as
\beq \label{eq:total-mass-Mg}
M_g = \frac{V_0^2 R_g}{A \, G} \, . 
\eeq

To check viability we investigate the idealized rotational 
velocity profile $V(r)$ of our own Milky Way galaxy shown in 
Figure \ref{fig:fig1} 
with core radius $R_c = 0.015$. 
From galactic rotation measurements, 
the parameters appropriate for the Milky Way galaxy are $R_c = 0.015$, 
$V_0 = 2.5 \times 10^5 (m/s)$, and  
$R_g = 10^5 (\mbox{light-years}) = 9.46 \times 10^{20} (m)$. 
In Table 1, the total galactic mass $M_g$ of the Milky Way galaxy is calculated for a wide range of bulge masses $M_b$ and bulge sizes $\beta$, and the corresponding computed values of galactic rotation parameter $A$.      
The total mass of the Milky Way galaxy is then determined  
from (\ref{eq:total-mass-Mg}) to be in the range 
\[ 
M_g = 2.8 - 3.2 \times 10^{11} (\mbox{solar-mass})
\, .
\]   
These values are in very good agreement with Milky Way star counts of 100 billion. 
In Table 1, note that a large increase in bulge mass $M_b$ only yields a small increase (10 percent) in the total galactic mass $M_g$ over the disk-only case (no bulge) ($M_b = 0$). However, these larger galactic masses may possibly be more compatible with reality since the galaxies also contain gases, dust, grains, lumps, planets and plasma, all in addition to stars. 
 
However, we emphasize the essential physics of galactic rotation is 
gravitationally controlled by the ordinary baryonic matter within 
thin galactic disks. The central bulge has minor effects because the comparatively small amount of matter in the outer regions of the disk are gravitationally more effective in controlling the rotational dynamics of the galactic periphery.     



\begin{table}
  \begin{center}
  \begin{tabular}{lccc}
    $M_b$ & $\beta$ & $A$ & $M_g$ (solar-mass) \\[3pt]
     0 & -- & 1.57 & 2.84 $\times 10^{11}$ \\
     0.1 & 500 & 1.57 & 2.84 $\times 10^{11}$ \\
     0.1 & 100 & 1.56 & 2.86 $\times 10^{11}$ \\
     0.15 & 200 & 1.56 & 2.86 $\times 10^{11}$ \\
     0.2 & 200 & 1.56 & 2.86 $\times 10^{11}$ \\
     0.25 & 100 & 1.54 & 2.89 $\times 10^{11}$ \\
     0.3 & 50 & 1.52 & 2.93 $\times 10^{11}$ \\
     0.4 & 20 & 1.47 & 3.03 $\times 10^{11}$ \\
     0.5 & 10 & 1.39 & 3.21 $\times 10^{11}$ \\
  \end{tabular}
  \caption{For the Milky Way galaxy, the total galactic mass $M_g$ (in units of solar-mass) is      
  calculated (\ref{eq:total-mass-Mg}) from data and various bulge parameters. From the measured 
  galactic rotation curve, 
  $R_c = 0.015$, $R_g = 9.46 \times 10^{20}$ (m), and $V_0 = 2.5 \times 10^5$ (m/s).
  A range of values of bulge parameters ($M_b$, $\beta$) are examined along with the corresponding   
  computed values of the galactic rotation parameter $A$. The computed total galactic mass $M_g$ 
  increases only 10 percent over the disk-only case (no bulge) ($M_b = 0$) in spite of large increases 
  of the examined bulge mass $M_b$.} 
  \end{center}
\end{table}

\section{Ordinary Baryonic Matter versus Dark Matter}
To theoretically describe the measured rotational velocity curves of spiral galaxies, there
are three very different approaches and conclusions. 

(1) Ordinary Baryonic Matter. 
We assume Newtonian gravity/dynamics and computationally solve for mass distributions that successfully duplicate the measured rotational velocities. These mass distributions decrease roughly exponentially from the galactic center in the central core, but then decrease more slowly (inversely with radius) towards the periphery. This decrease is slower than the measured light distribution. Thus there is ordinary baryonic matter within the galactic disk distributed towards the cooler periphery with lower emissivity/opacity and therefore darker. Our view is consistent with edge-views of galaxies which exhibit a dark disk line against a much brighter galactic halo. There are no mysteries in this rational scenario based on verified physics 
(Refs.\cite{Marmet}-\cite{BalasinG}). 
 
(2) Dark Matter.  
By contrast, others inaccurately assume the galactic mass distributions follow the measured light distributions (approximately exponential), and then the measured rotational velocity curves are not  duplicated. But this assumption of a simple direct relationship between light intensity and mass is very inaccurate because it is not based on sound physical principles. This so-called Mass/Light ratio is inaccurate since both the temperature and opacity/emissivity are important but ignored variables. These deficiencies are clear from edge-on views of spiral galaxies where a dark galactic line is obvious against a bright galactic background. revealing the substantial radial temperature gradient across the galaxy. There is no simple direct relationship between mass and light, and such an assumption is grossly over-simplified (Refs.\cite{Freeman}-\cite{deJong2}). 

With this inaccurate assumption, the discrepancy between measured and calculated velocity profiles are particularly severe beyond the galactic core. To alleviate this discrepancy, speculations are invoked re ``massive peripheral spherical halos of mysterious Dark Matter'' But no significant matter has been detected in this untenable unstable gravitational halo distribution. This speculated Dark Matter is ``mysterious'' since it does not interact with electromagnetic fields (light) nor ordinary matter except through gravity. This Dark Matter must have other abnormal (non-baryonic) properties to maintain its peripheral spherical shape against the galactic rotation and gravitational attraction of ordinary matter. Many unverified ``mysteries'' are invoked as solutions to real physical phenomena  
(Refs.\cite{Freeman}-\cite{deJong2}). 

(3) Modified Gravity.  
Possible deviations from Newtonian gravtiy/dynamics have been proposed, but there is no independent experimental evidence of such deviations. Our use of Newtonian gravity/dynamics with sound computational techniques has proven successful to explain the observed flat rotation curves
(Refs.\cite{Milogram}).   

Conclusion.  
We conclude our approach utilizing Newtonian gravity/dynamics and computationally solving for the ordinary baryonic mass distributions within the galactic disk simulates reality and agrees with data.

\section{Nature of Ordinary Baryonic Matter} 

Our approach yields higher total galactic masses in agreement with star counts. Concurrently, our mass density distributions also yield more mass distributed out towards the cooler galactic periphery. We wish to conjecture about the nature of some of this ordinary baryonic matter which appears dark. In addition to stars, this material is some combination of dust, grains, lumps, planets, plasma and gases.  

Consider hydrogen gas. Since the temperature is higher in the galactic core and cooler towards the galactic periphery, we expect more ionized hydrogen (plasma) in the hotter core, then more atomic hydrogen away from the core, and finally more molecular hydrogen out towards the cooler periphery. This molecular hydrogen is often ignored, although measurements have revealed its presence (Ref.\cite{H2Measure}). However, quantitative estimates of its density vary widely. We note that molecular hydrogen is a naturally dark material since it has very low emission and absorption coefficients due to its high molecular structural symmetry. Combined with its presence towards the cooler periphery, we qualitatively believe this molecular hydrogen  (Refs.\cite{PaulMarmet}\cite{GalloH2Marmet}) is one component of the ordinary baryonic dark matter in the disk periphery we have found from analysis of galactic rotation profiles.

\section{Limitations and Strengths of Disk and Bulge+Disk Models}
Our Disk and Bulge+Disk gravitational models do not address many important features 
such as spiral structure, plasma effects, galactic formation, 
galactic evolution, galactic jets, black holes, relativistic effects, 
galactic clusters, etc.. 

It is well known (Ref.\cite{BT}) that the internal gravitational behavior 
of a thin disk is much different than a sphere.  
This distinctly different behavior of disks enables our models  
to describe the rotational dynamics of mature spiral galaxies, 
and their total galactic masses, even with the addition of substantial central spherical bulges. The reason is that matter in the outer disk periphery is gravitationally closer and stronger in controlling the rotation in those outer regions, even compared with the substantial central spherical bulges which are more distant and gravitationally weaker.   

Our gravitational models have finite radial extent. 
Beyond the galactic radius, we assume the density has dropped to 
the inter-galactic level, which is approximately spherically symmetric 
and thus no longer affects the galactic dynamics. 
We mention this because some others (Refs.\cite{BT}\cite{Freeman}),
have taken 
the relevant integrals to infinity, which we think is inappropriate. 

In our approaches, we balance the gravitational forces against 
the centrifugal forces at each and every point within the disk. 
Thus, our solutions for the mass distributions and total galactic mass 
satisfy the rotational velocity measurements and ensure stability within the same context 
as similar calculations for our Solar System and Earth satellites. 
Some previous authors obtain solutions that are not gravitationally 
stable because they obtain incorrect mass distributions and incorrect 
galactic masses and do not satisfy the measured rotational profiles. 
Thus, their solutions are unstable, whereas our solutions are stable 
within the Newtonian context.  

Plasma effects are certainly active in the formation and 
evolution of galaxies from the original hot 
plasma (Refs.\cite{Peratt1}-\cite{Peratt3}). However, for mature 
spiral galaxies, the free plasma density has 
dropped to levels sufficiently low that plasma does not affect 
the predominantly gravitational galactic dynamics. 
This is evidenced in our own Solar System in which gravitational dynamics 
dominate even with the observed effects of solar wind, coronal mass ejections, auroras,  
comet tails, etc. The plasma in our Sun is stabilized by gravitational forces, 
even though plasma effects are very active within the Sun itself. 
Since our Solar System is approximately 1/3 distance from the center of our 
Milky Way galaxy, we have our Solar System evidence for 
the dominance of gravitational forces within our own Milky Way galaxy, 
at least at this radial distance and beyond to the periphery. We expect plasma phenomena to be more active in the hotter central galactic core.       

Summarizing, both our Disk and Bulge+Disk models are sufficient 
to describe the rotational dynamics of mature spiral galaxies 
and their total galactic masses. Disk models with an additional central Bulge yield higher total galactic masses. 

\section{Conclusions} 

Even with the addition of substantial central galactic bulges, all the critical essential features of our thin disk gravitational models are preserved. 
(1) Balancing Newtonian gravitational and centrifugal forces 
at every point within the disk yields computed radial mass distributions that describe the measured rotation velocity profiles of mature spiral galaxies successfully. 
(2) There is no need for gravity deviations or ``massive peripheral spherical halos of mysterious Dark Matter''. 
(3) The calculated total galactic masses are in good agreement with star count data. 
(4) The addition of central bulges increases the calculated total galactic masses, possibly more consistent with the presence of galactic gases, dust, grains, lumps, planets and plasma in addition to stars. 
(5) Compared with the light distribution, our mass distributions within the disk are larger out toward the galactic periphery which is cooler with lower opactiy/emissivity (and thus darker). This is apparent from edge-on views of galaxies which display a dark disk-line against a much brighter galactic halo.  

Most previous research assumes a galactic density decreasing exponentially with radius out to the galactic periphery, analogous to the  measured light distribution. But this assumption (by others) is inaccurate since both the temperature and opacity/emissity are important but ignored variables. There is no simple relationship between mass and light. These prior models do NOT describe the measured velocity profiles, and speculations are invoked re halos of mysterious Dark Matter or gravitational deviations to compensate. The Dark Matter must have ``mysterious'' (non-baryonic) properties because there is no evidence of its existence and it is not responding to gravitational, centrifugal and electromagnetic forces in any known manner. 
By contrast, our results indicate no massive peripheral spherical halos of mysterious Dark Matter and no deviations from simple gravity. Our total galactic mass determinations are also in reasonable agreement with data.   

The controversy is summarized as follows. 

We believe there is ordinary baryonic matter within the galactic disc distributed more towards the galactic periphery which is cooler with lower opacity/emissivity (and therefore darker). 

Others believe there are massive peripheral spherical halos of mysterious Dark Matter surrounding the galaxies.

\section{Acknowledgements}
We gratefully acknowledge Louis Marmet, Ken Nicholson and Michel Mizony whose intuition and computational techniques convinced us that galactic rotation could be described by suitable mass distributions of ordinary baryonic matter within galactic disks. 
Anthony Peratt originally sparked our interest with his plasma dynamical calculations re the formation and evolution of galaxies. Ari Brynjolfsson has energetically supported our efforts.


\begin{thebibliography}{99}


\bibitem{Rubin1} 
Rubin and Ford, 
Astrophysical Journal, Vol.159, pp.379, 1970. 

\bibitem{RobertsW} 
Roberts and Whitehurst, 
Astrophysical Journal, Vol.201, pp.327, 1975. 

\bibitem{BahcallC} 
Bahcall and Casertano, 
Astrophysical Journal, vol. 293, L7, 1985. 

\bibitem{BursteinRubin} 
Burstein and Rubin, 
Astrophysical Journal, vol. 297, pg. 423, 1985. 

\bibitem{PersicS} 
Persic and Salucci, 
Astrophysical Journal Supplement, vol. 99, pg. 501, 1995. 
 
\bibitem{deBlockMH} 
deBlock, McGaugh and vanderHulst, MNRAS, Vol.274, pg.235, 1995. 

xxxxxxxxxxxxxxxxxxxxxxxxxxxxxxxxxxxxxxxxxxxxx

\bibitem{BT}
Binney, J. and Tremaine, S. 
``Galactic Dynamics'', 
Princeton University Press, Princeton, 1987.

xxxxxxxxxxxxxxxxxxxxxxxxxxxxxxxxxxxxxxxxxxxxxxxx

\bibitem{FengGallo1} 
Feng, J. Q. and Gallo, C. F., 
``Galactic Rotation Described with Thin-Disk Gravitational Model'', 
arXiv:0803.0556v1 [astro-ph], 4 Mar 2008. 

\bibitem{FengGallo2} 
Feng, J. Q. and Gallo, C. F., 
``Galactic Rotation Described with Various Thin-Disk Gravitational Models'', 
arXiv:0804.0217v1 [astro-ph], 31 Mar 2008. 

xxxxxxxxxxxxxxxxxxxxxxxxxxxxxxxxxxxxxxxxxxxxxx

\bibitem{GalloVortex}
Gallo, C. F., 
``Spiral Galaxy Model with Axial Plasma/Gas Vortex: A Possible Suggestion''.  
APS Conference - April 14-17, 2007, Jacksonville FL,  
Bull Am Phys Soc, Vol.52, No.3, April 14-17, 2007.  
Poster S1-9. p 170. 

xxxxxxxxxxxxxxxxxxxxxxxxxxxxxxxxxxxxxxxxxxxxxxxx

\bibitem{PressTVF}
Press, W. H., Teukolsky, S. A., Vetterling, W. T., and Flannery, B. P., 
``Numerical Recipes'', Cambridge University Press, Cambridge, 1988.

xxxxxxxxxxxxxxxxxxxxxxxxxxxxxxxxxxxxxxxxxxxxxxxxx


\bibitem{Marmet}
Marmet, Louis, 
``Rotation Dynamics of a Galaxy with a Double Mass Distribution'', 
http://www.marmet.org/cosmology/galaxy rotation/dynamics.pdf.

\bibitem{nicholson}
Nicholson, Kenneth F., 
``Galactic mass distribution without dark matter or modified Newtonian mechanics'',   
arXiv:astro-ph/0309762 v2 [pdf]. 

``Errors in equations for galaxy rotation speeds'',   
arXiv:astro-ph/0309823 [pdf]. 

``Galaxy Mass Distributions from Rotation Speeds by Closed-Loop Convergence'',  
arXiv:astro-ph/0303135 [pdf]. 

``Galaxy mass distributions for some extreme orbital-speed Profiles'',  
arXiv:astro-ph/0101401 [pdf]. 

``Disk-galaxy density distribution from orbital speeds using Newton's law'',  
arXiv:astro-ph/0006330 v1.1 [pdf]. 

``Disk-galaxy density distribution from orbital speeds using Newton's law'',   
arXiv:astro-ph/0006140 [pdf]. 

\bibitem{meraMB}
Mera, Mizony and Baillon, 
``Disk Surface Density Profile of Spiral Galaxies and Maximal Disks'',  
Astronomy and Astrophysics.   
Preprint Lyon (1997).
http://math.univ-lyon1.fr/mizony/michel/pdfch8bis.pdf
``Substellar Mass Function and Maximum Baryonic Mass in the Halo of the Galaxy'', 
Mera et al, 1996 Europhys. Lett. 33, 327-332. 

\bibitem{mizony1}
Mizony, M.,   
``Flatness of the Rotation Curves of the Galaxies: 
Exit the Recourse to a Massive Halo''. 
English Summary of Above Publication by Mera, Mizony and Baillon.  
http://igd.univ-lyon.fr/mizony/rotation.html. 
APS Conference - April 14-17, 2007, Jacksonville FL.   
Bull Am Phys Soc, vol 52, no 3, April 14-17, 2007. 
Y11-4, p.208.  
http://meetings.aps.org/Meeting/APR07/Event/65677

\bibitem{BaillonMizony} 
Baillon and Mizony, 
Poster: Title: ``On the Hidden Mass'', 
Symposium UAI number 172: Dynamics; Ephemerides and Astrometry of the Solar
System, Paris, 3-8 July 1995. 
(Abstract in
http://www.imcce.fr/page.phpnav-fr/publications/colloque\_sympo/SYMP172/symp.php)

\bibitem{Mizony2}
Mizony, 
``La relativit\'e g\'en\'erale aujourd\'hui ou l\'observateur oubli\'e'',
Editions Al\'eas, Juin 2003, p.276, Chap. 9.

\bibitem{mizonyLR}
Mizony and Lachieze-Rey
``Cosmological Effects in the Local Static Frame'',  
arXiv.gr-qc/0412084 v1 17Dec2004.  

\bibitem{meraCS}
Mera, Chabrier and Schaeffer
``Towards a Consistent Model of the Galaxy: II Derivation of the Model'',  
Astronomy and Astrophysics, vol. 330, pp. 953-962, 1998.  

\bibitem{FengG}
Feng, James Q. and Gallo, C. F., 
``Rotation of Spiral Galaxies Described with a Simple Disc Gravitational Model''.   
APS Conference - April 14-17, 2007, Jacksonville FL,  
Bull Am Phys Soc, Vol.52, No.3, April 14-17, 2007.  
E12-4. p 62. 

\bibitem{GalloF1}
Gallo, C. F. and Feng, James Q., 
``Rotation of Spiral Galaxies Described with Various Sphere+Disc Gravitational Models''. 
APS Conference - April 14-17, 2007, Jacksonville FL,  
Bull Am Phys Soc, Vol.52, No.3, April 14-17, 2007.  
E12-5. p 62.  

\bibitem{GalloF2}
Gallo, C. F. and Feng, James Q., 
``Ordinary Dark Matter versus Mysterious Dark Matter in Galactic Rotation''.  
APS Conference - April 12-15, 2008, St Louis MO,  
Bull Am Phys Soc,  Vol.53, No.3?, April 12-15, 2008.  
H8-6. p.. 

\bibitem{GalloFH22}  
Gallo and Feng, 
``Ordinary Dark Matter versus Mysterious Dark Matter in Galactic Rotation''.
Crisis in Cosmology Conference: CCC-2.  
Sept 7 -11, 2008 - Port Angeles  WA.  

\bibitem{Hure}
Hure,  
``Solutions of the Axi-Symmetric Poisson Equation from Elliptic Integrals''.
``I. Numerical Splitting Methods'',  
Astronomy and Astrophysics, vol 434, pp. 1-15 (2005). 

\bibitem{PierensH}
Pierens and Hure, 
``Solutions of the Axi-Symmetric Poisson Equation from Elliptic Integrals''.
``II. Semi-Analytical Approach'',  
Astronomy and Astrophysics, vol 434, pp. 17-23 (2005). 
(astro-ph/0312529). 

\bibitem{pronko}
Pronko, G., 
``Flat Rotation Curves Without Dark Matter'', 
http://arXiv:astro-ph/0611303v1.               

\bibitem{FuchsBMZ}
B. Fuchs, A. Bohm, C. Mollenho and B. L. Ziegler
``Quantitative interpretation of the rotation curves of 
spiral galaxies at redshifts z = 0.7 and z = 1'',  
http://arxiv:astro-ph/0408072.

\bibitem{banhatti}
Banhatti, Dilip,  
``Disk galaxy rotation curves and dark matter distribution'', 
arXiv:astro-ph/0703430v2.  

\bibitem{RevazPCB}
Revaz, Pfenniger, Combes and Bournaud, 
``Simulations of galactic disks including a dark baryonic component'', 
arXiv:0801.1180, 8 Jan 2008. 

\bibitem{Mendez}  
R. Mendez, A. Riffeser, R.-P. Kudritzki, M. Matthias, K. Freeman, M. Arnaboldi, M. Capaccioli, and
O. Gerhard, 
``Detection, Photometry and Slitless Radial Velocities of 535 Planetary Nebulae in the Flattened
Elliptical Galaxy NGC 4697'', 
Astrophysical Journal , 2001. 
arXiv:astro-ph/0109075v1, Submitted to the Astrophysical Journal.

\bibitem{delRio} 
M. S. del Rio, 
``NGC 404, a Galaxy without Dark Matter'', 
Revista Mexicana de Astronomia y Astrofisica Serie de Conferencias, 
Aguilar and Carramiana, ed., p. 121, 
IX Latin American Regional IAU Meeting, 07 2001. 
Held in Tonantzintla, Mexico, Nov 9-13, 1998.

\bibitem{Bertin}
G. Bertin, F. Bertola, L. M. Buson, I. J. Danzinger, H. Dejonghe, E. M. Sadler, R. P. Saglia, P. T. de Zeeuw, and W. W. Zeilinger, 
``A Search for Dark Matter in Elliptical Galaxies: Radially Extended Spectroscopic
Observations for Six Objects'', 
Astronomy and Astrophysics 292(2), pp.381–391, 1994.

\bibitem{cooperstock}
Cooperstock, F. and Tieu, S., 
``Galactic Dynamics via General Relativity: 
A Compilation and New Developments'',  
http://arxiv:0610370, astro-ph/0507619, astro-ph/0512048.  
Mod Phys Lett A 21, 2133 (2006).

\bibitem{BalasinG}
Balasin and Grumiller.    
``Significant Reduction of Galactic Dark Matter by General Relativity'', 
arXiv:astro-ph/0602519v2.   

xxxxxxxxxxxxxxxxxxxxxxxxxxxxxxxxxxxxxxxxxxxxxxxxxxx

\bibitem{Freeman} 
Freeman, K.C. 
``On the Disks of Spiral and SO Galaxies'', 
Astrophysical Journal, vol. 160, pg. 811, June 1970.

\bibitem{Mestel} 
Mestel, 
MNRAS, vol. 126, pg. 553, 1963.  

xxxxxxxxxxxxxxxxxxxxxxxxxxxxxxxxxxxxxxx

\bibitem{Bennett} 
Bennett, Donahue, Schneider, and Voit, 
``Cosmic Perspective: Stars, Galaxies, and Cosmology'', Addison and Wesley, 2007.  

xxxxxxxxxxxxxxxxxxxxxxxxxxxxxxxxxxxxxxxxxxxxxxx


\bibitem{OstrikerPY} 
Ostriker, Peebles and Yahil, 
Astrophysical Journal Letters, Vol.193, pp.L1, 1974. 

\bibitem{FaberG}  
Faber and Gallagher, 
Astronomy and Astrophysics, Vol.17, pp.135, 1979. 

\bibitem{Rubin2}
Rubin, Vera,  
``Galaxy Dynamics and the Mass Density of the Universe,'' 
Proc. Natl. Acad. Sci. USA 90, pp. 4814–4821, June 1993.

\bibitem{Rubin3}
Rubin, Vera, 
``Seeing Dark Matter in the Andromeda Galaxy'', 
Physics Today, vol.59, No.12, pp.8-9, Dec 2006. 
``A Two-Way Galaxy'', 
Physics Today, vol.60, No.9, pp.8-9, Sept 2007, 

\bibitem{Albada}
van Albada, Bahcall, Begeman, and Sanscisi, 
``Distribution of Dark Matter in the Spiral Galaxy NGC 3198'', 
Astrophysical Journal 295, pp. 305–13, August 15 1985.

\bibitem{DalcantonSpergelSummers} 
Dalcanton, Spergel and Summers, 
``The Formation of Disk Galaxies'', 
Astrophysical Journal, vol. 482, pp. 659-676, June 20, 1997. 

\bibitem{McGaugh1}
McGaugh, Stacy, 
``Observational Constraints on the Acceleration Discrepancy Problem'', 
arXiv:astro-ph/0606351. 

\bibitem{Burstein}
Burstein, 
``The Distribution Of Mass In Sc Galaxies'', 
Astrophysical Journal 253, pp. 70–85, February 1 1982.


\bibitem{Alves}
Alves and Nelson, 
``The Rotation Curve of the Large Magellanic Cloud and the Implications for Microlensing'',  
arXiv: astro-ph/0006018, 6/1/2000. 

\bibitem{Bok}
Bok, R.J., 
``The Milky Way Galaxy'', 
Scientific American, March 1981. 

\bibitem{Kim}
Kim, S. et al, 
``An H1 aperture synthesis mosaic of the Large Magellanic Cloud'', 
ApJ 503,674K 08/1998. 

\bibitem{Kunkel} 
Kunkel, W.E. et al, 
``The Dynamics of the Large Magellanic Cloud Periphery: Mass Limit and Polar Ring'',
16 ApJ 488 10/20/1997. 

\bibitem{Noodemeir}
Noodermeir et al, 
``Rotation Curves and Dark Matter in Early Type Disk Galaxies'',
PoS(BDMH2004)068, 10/9/2004. 

\bibitem{Weldrake}  
Weldrake, D.T.F. et al, 
``A High-resolution Rotation Curve of NGC 6822: A Test Case for Cold Dark Matter'', 
arXiv: astro-ph/0210568, 10/25/2002

\bibitem{Fall,E} 
Fall and Efstathiou, 
MNRAS, vol. 93, pg. 189, 1980. 

\bibitem{Persic} 
Persic, Salucci and Stel, 
``The Universal Rotation Curve of Spiral Galaxies - I. The Dark Matter Connection'', 
MNRAS Vol.281, pp. 27-47 (1996). 

\bibitem{Kruit} 
van der Kruit, 
Astronomy and Astrophysics, vol. 173, pg. 59, 1987. 

xxxxxxxxxxxxxxxxxxxxxxxxxxxxxxxxxxxxxxxxxxxxxxxxxxxxxx

\bibitem{deJong1} 
deJong,  
Astronomy and Astrophysics, vol. 118, p.557, 1996.  
 
\bibitem{DaviesPD} 
Davies, Phillips and Disney, 
MNRAS, vol.231, pg.69P, 1988; vol. 244, pg. 385, 1990. 
 
\bibitem{McGaughB} 
McGaugh and Bothum, 
Astrophysical Journal, vol.107, pg.530, 1994. 

\bibitem{McGaughSB} 
McGaugh, Schombert and Bothum, 
Astrophysical Journal, vol.109, pg.2019, 1995. 

\bibitem{DalcantonSGSS} 
Dalcanton, Spergel, Gunn, Schmidt and Schneider, vol.xx, pg.xx, 1997.  

\bibitem{deJong2} 
deJong and van der Kruit, 
Astronomy and Astrophysics, vol. 106, p.451, 1994.  

xxxxxxxxxxxxxxxxxxxxxxxxxxxxxxxxxxxxxxxxxxxxxxxxxxxxxxx


\bibitem{Milogram} 
Milgrom, 
``A Modification of Newtonian Dynamics as a Possible Alternative to the Hidden Mass
Hypothesis'', 
Astrophysical Journal vol.270, p. 365, 1983.
Milgrom, M., 
``A Modification of Newtonian Dynamics: Implications for Galaxies'', 
ApJ 270: 371-388, 7/15/1983

xxxxxxxxxxxxxxxxxxxxxxxxxxxxxxxxxxxxxxxxxxxxxxxxx


\bibitem{H2Measure}
E. A. Valentijn and P. P. van der Werf, ``First Extragalactic Direct Detection of Large-Scale Molecular Hydrogen in the Disk of NGC 8911,'' The Astrophysical Journal 522, pp. L29-L33, September 1 1999. 
Note that molecular hydrogen seems to outweigh atomic hydrogen by factor 5-15. 
This factor matches the mass required to resolve the problem of the missing matter of spiral galaxies within the disk.

\bibitem{PaulMarmet}
P. Marmet, ``A New Non-Doppler Effect,'' Physics Essays, Vol.1, p. 24 (1988). 
Published by: 21st Century, Science and Technology, Washington, D.C. 20041. Vol.3, No.2, Spring 1990, P. 52-59.

Paul Marmet, 21st Century Science and Technology, Spring 2000, Pages 5-7.  

P. Marmet and G. Reber, 1989. ``Cosmic Matter and the Non-Expanding Universe'', IEEE Transactions on Plasma Science, Vol. 17, No. 2 (Dec.), pp. 264-269.

P. Marmet, 1990. ``Big Bang Cosmology Meets an Astronomical Death'', 21st Century Science and Technology, Vol. 3, No. 2 (Spring), pp. 52-59. 

P. Marmet, 1990. ``The Deceptive Illusion of the Big Bang Cosmology'', Physics in Canada, Vol. 46, No.5, pp. 97-101. 

\bibitem{GalloH2Marmet}
Gallo, ``Modeling the Rotational Dynamics of Spiral Galaxies with Plasmas, Molecular Hydrogen, and Numerical Mass Distributions (Elliptical + Disk)''. 
Abstract Poster D1-93. p. 65.   
APS Conference - April 22-25, 2006 - Dallas TX. 
Bull Am Phys Soc Vol. 51, No.2, April 22-25, 2006.
 
xxxxxxxxxxxxxxxxxxxxxxxxxxxxxxxxxxxxxxxxxxxxxxxxxx

\bibitem{Peratt1} 
Peratt, Anthony, 
"Evolution of the Plasma Universe: I. Double Radio Galaxies, Quasars, and Extragalactic Jets", 
IEEE Trans. Plasma Sci. Vol.PS-14, No.6, pp.639-660, December 1986.

\bibitem{Peratt2} 
Peratt, Anthony 
"Evolution of the Plasma Universe: II. The Formation of Systems of Galaxies", 
IEEE Trans. Plasma Sci. Vol.PS-14, No.6, pp.763-778, December 1986.

\bibitem{Peratt3}
Peratt, Anthony. 
"Physics of the Plasma Universe" (Springer-Verlag, 1992). 

xxxxxxxxxxxxxxxxxxxxxxxxxxxxxxxxxxxxxxxxxx






\end{thebibliography}
\end{document}